\documentclass[amsmath,showpacs,aps,twocolumn,prb,floatfix]{revtex4-1}
\usepackage{bm}
\usepackage{graphics}
\usepackage{graphicx}
\usepackage{amsthm}
\usepackage{amsmath}
\usepackage{wasysym}
\usepackage{amssymb}
\usepackage{enumerate}
\usepackage[dvips]{epsfig}

\begin{document}
\title{Exact Chiral Spin Liquids and Mean-Field Perturbations of Gamma Matrix Models on the Ruby Lattice}
\author{Seth Whitsitt}
\author{Victor Chua}
\author{Gregory A. Fiete}
\affiliation{Department of Physics, The University of Texas at Austin, Austin, TX 78712, USA}
\begin{abstract}
We theoretically study an exactly solvable Gamma matrix generalization of the Kitaev spin model on the ruby lattice, which is a honeycomb lattice with ``expanded" vertices and links.  We find this model displays an exceptionally rich phase diagram that includes: (i) gapless phases with stable spin fermi surfaces, (ii) gapless phases with low-energy Dirac cones and quadratic band touching points, and (iii) gapped phases with finite Chern numbers possessing the values $\pm4, \pm 3,\pm 2$ and $\pm 1$. The model is then generalized to include Ising-like interactions that break the exact solvability of the model in a controlled manner. When these terms are dominant, they lead to a trivial Ising ordered phase which is shown to be adiabatically connected to a large coupling limit of the exactly solvable phase. In the limit when these interactions are weak, we treat them within mean-field theory and present the resulting phase diagrams. We discuss the nature of the transitions between various phases. Our results highlight the richness of possible ground states in closely related magnetic systems. \end{abstract}

\pacs{71.10.Jm,71.10.Kt,75.50.Mm}


\maketitle

\section{Introduction}
The study of zero-temperature properties of ``frustrated" magnetic systems has been reinvigorated in recent years with the development of powerful numerical methods,\cite{White:sci11,Evenbly:prl10,Singh:prb07} the discovery of new classes of exactly solvable models,\cite{Kitaev:ap03,Kitaev:ap06} and the realization that intriguing topological ground states could occur.\cite{Balents:nat10,Wen}  In particular, there is a large class of exactly solvable quantum spin models known as Kitaev models\cite{Kitaev:ap06} which have revitalized the study of exactly solvable spin-liquid systems. The appeal of this class of models lies in the relative ease in which Kitaev's original version can be generalized, spawning many variants,\cite{Yao:prl07,Yang:prb07,Yu:npb08,Yao:prl09,Wu:prb09,Baskaran:09,Nussinov:prb09,Chern:prb10,Ryu:prb09,Mandal:prb09,Yao:prl11,Kells:prb10} and its possession of many non-trivial properties\cite{Tikhonov:prl10,Vidal:prb08,Yao:prl10,Chung:prb10,Lee:prl07,Baskaran:prl07,Chen:jpa08,Lahtinen:njp11,Chua:prb11,Kells:njp11}
even in the presence of disorder.\cite{Willans:prl10,Dhochak:prl10,Chua_dis:prb11} Moreover, the exact solution of Kitaev models requires no more effort than solving the problem of non-interacting particles moving in a static background magnetic field. Nonetheless, the exact eigenstates of all Kitaev models are non-trivial entangled many-body wavefunctions.\cite{Yao:prl10,Chung:prb10} More specifically, their respective ground states are examples of quantum spin-liquids, which are insulating quantum states of matter exhibiting no conventional long range magnetic order at zero temperature. Although no physical example of these models has been established in nature, there have been several experimental proposals for realizing them.\cite{Duan:prl03,Micheli:natp06,Jackeli:prl09,Roustekoski:prl09,Chaloupka:prl10,You:prb10,Kimchi:prl11} Nevertheless, the importance of this entire class of models lies in providing a strong case - by proof of principle - for the existence of exotic emergent phenomena in many-body quantum phases of matter such as quantum spin-liquids, as well as their utility as model systems for the study of non-trivial, non-perturbative emergent quantum phenomena in general.

In this paper we study a new exactly solvable Kitaev model and examine its response to the inclusion of interactions that destroy the exact solvability. This will motivate us to analyze the order that is favored by these new interactions and their effects on the exactly solvable ground state. The model is a Gamma matrix model (GMM) extension \cite{Yao:prl09,Wu:prb09,Ryu:prb09} on the two dimensional ruby lattice. Previous studies on the ruby lattice have yielded a topological insulator, \cite{Hu:prb11} fractional quantum anomalous Hall states,\cite{Wu:prb12} and also topological anyons.\cite{Bombin:prb09} Here we show that a GMM on the ruby lattice realizes a quantum spin liquid ground state with an unusually rich phase diagram. The model is then generalized by the inclusion of Ising-like interactions that spoil the exact solvability of the model. We explore the interplay between these interactions and the many ground state spin liquid phases.

The paper is organized as follows. In section II, we introduce the Hamiltonian and discuss its solution by mapping onto a system of non-interacting Majoranas moving in an emergent $\mathbb{Z}_2$ gauge field. In section III, we discuss the various ground state phases that the model realizes and present a phase diagram. In section IV we examine the Ising order in one of the gamma matrix operators which may be induced by including Ising-like interactions. In section V we treat these extra interactions within the mean field approximation and discuss their effects on the phase diagram. We then end with conclusions in section VI.

\section{Gamma-matrix Model and Hamiltonian}
\label{sec:Gamma}

\begin{figure}[h]
\includegraphics[width=8cm]{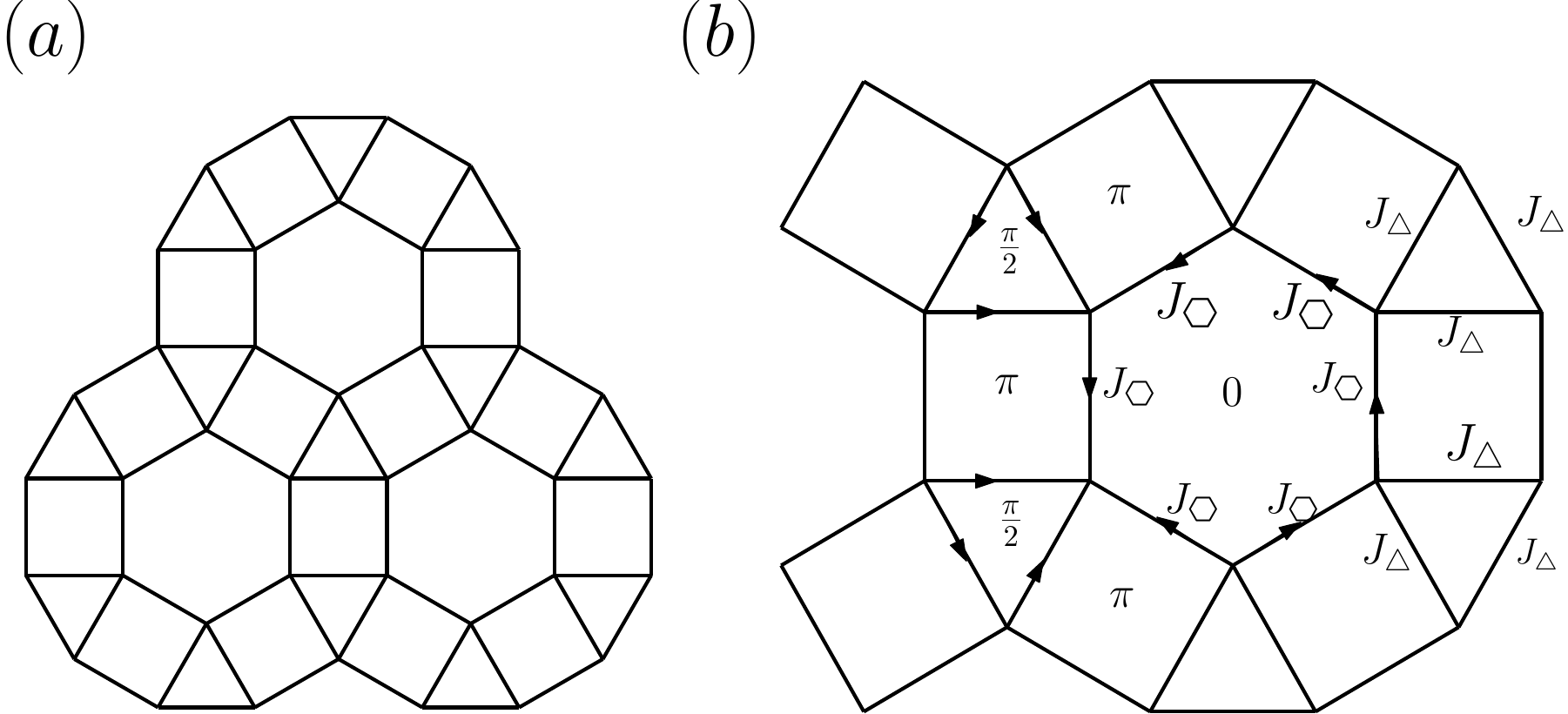}
\caption{(a) The ruby lattice.  The ruby lattice can be viewed as an ``expanded" honeycomb lattice with triangles replacing the vertices and squares replacing the bonds.  (b) The couplings in our Hamiltonian Eq.\eqref{eqn:Hamiltonian} and ground state flux configuration of the $\mathbb{Z}_2$ gauge field. Note the $J'$ couplings (not shown) are also defined on the same links as the $J$'s. The arrows indicate the choice of gauge used for the $u_{ij}$ fields in \eqref{eqn:H_uij}.} \label{fig:lattice_ruby}
\end{figure}

We consider spin-$3/2$ moments located on the sites of a ruby lattice, shown in Fig.\ref{fig:lattice_ruby}(a). Alternatively, one can think of each site as being occupied by a single spin-$1/2$ which has an additional orbital degree of freedom. At each site we introduce five 4$\times$4 Hermitian Gamma matrices which satisfy a Clifford algebra locally $\{\Gamma^a_i,\Gamma^b_i\}=2\delta_{ab}$, with $a,b=1,\dots 5$\cite{Yao:prl09,Wu:prb09,Ryu:prb09}. These matrices also have a representation in terms of bilinears of $SU(2)$ spin-3/2 operators on each site $i$ as follows,
\begin{eqnarray}\label{eqn:spin32}
&&\Gamma_i^1 = \frac{1}{\sqrt{3}} \{ S_i^y , S_i^z \}, \quad \Gamma_i^2 = \frac{1}{\sqrt{3}} \{ S_i^x,S_i^z \}, \nonumber \\
&&\Gamma_i^3 = \frac{1}{\sqrt{3}} \{ S_i^x , S_i^y \}, \quad \Gamma_i^4 = \frac{1}{\sqrt{3}} [(S_i^x)^2 - (S_i^y)^2] ,\nonumber \\
&&\Gamma_i^5 = (S_i^z)^2 - \frac{5}{4}.\quad\quad\quad {}
\end{eqnarray}
The Gamma matrices may also be expressed as tensor products of Pauli spin-1/2 matrices $\{\sigma^\mu,\tau^\nu\}$ on the site $i$,
\begin{eqnarray}\label{eqn:spinorbit}
&&\Gamma_i^1 = \sigma_i^z\otimes \tau_i^y, \quad \Gamma_i^2 = \sigma_i^z\otimes\tau_i^x, \nonumber \\
&&\Gamma_i^3 = \sigma_i^y\otimes\tau_i^0, \quad \Gamma_i^4 = \sigma_i^x\otimes\tau_i^0, \nonumber \\
&&\Gamma_i^5 = \sigma_i^z\otimes\tau_i^z,
\end{eqnarray}
where $\sigma_i^0=\tau_i^0=1$, the 2 x 2 identity matrix. In the spin-orbit interpretation \eqref{eqn:spinorbit}, $\sigma$ acts on the real spin degree of freedom and $\tau$ on the orbital degree of freedom.  Thus, one can view the Gamma matrix model we study below as applying to both a spin-3/2 model and a two-orbital spin-1/2 model.  In either case, there are 4 physical states per lattice site.

The Hamiltonian we study is given by
\begin{eqnarray}\label{eqn:Hamiltonian}
\mathcal{H}\;&&= \mathcal{H}_0 + \mathcal{H}_5, \nonumber \\
\mathcal{H}_0 &&= \quad J_{\triangle} \sum_{\langle ij \rangle \in \triangle}  \Gamma^1_i \Gamma^2_j \ + \ J'_{\triangle} \sum_{\langle ij \rangle \in \triangle}  \Gamma^{15}_i \Gamma^{25}_j \nonumber \\
&&\quad +\ J_{\hexagon}  \sum_{\langle ij \rangle \in \hexagon} \Gamma^3_i \Gamma^4_j \ + \ J'_{\hexagon} \sum_{\langle ij \rangle \in \hexagon}  \Gamma^{35}_i \Gamma^{45}_j, \nonumber \\ 
\mathcal{H}_5&&=-\ J_5 \sum_i \Gamma^5_i,
\end{eqnarray}
where we define the operators $\Gamma_i^{ab} = [\Gamma_i^a,\Gamma_i^b]/(2i)$ in terms of the $\Gamma_i^a$ given in either Eq.\eqref{eqn:spin32} or Eq.\eqref{eqn:spinorbit}.  The couplings $J_\triangle,J'_{\triangle},J_{\hexagon}$ and $J'_{\hexagon}$ are taken to be positive and are link dependent as shown in Fig.\ref{fig:lattice_ruby}(b). For generic couplings, the Hamiltonian has translational and six-fold rotational lattice symmetry. With the interpretation in terms of spin-3/2 moments, it has global Ising spin symmetry under 180$^\circ$ rotations about the $z$-axis, and time-reversal symmetry (TRS), although TRS will be spontaneously broken in the ground state as we will show below.

While the model \eqref{eqn:Hamiltonian} has unusual spin symmetries in terms of the underling spin and orbital degrees of freedom, its structure allows an exact solution.  A key ingredient for the exact solvability of the model is the existence of conserved operators on every plaquette of the lattice which commute with the Hamiltonian.\cite{Kitaev:ap06}  These operators are the three ($\hat{W}_\triangle$), four ($\hat{W}_\square$), and six point ($\hat{W}_{\hexagon}$) operators defined  on the triangles, squares, and hexagons of the lattice (as their subscripts suggest). Explicitly, they are given by $\hat{W}_{\triangle} = \Gamma^{12}_i \Gamma^{12}_j \Gamma^{12}_k,$ $\hat{W}_{\hexagon} = \Gamma^{34}_i \Gamma^{34}_j \Gamma^{34}_k \Gamma^{34}_l \Gamma^{34}_m \Gamma^{34}_n,$ and $\hat{W}_{\square} = \Gamma^{23}_i \Gamma^{14}_j \Gamma^{23}_k \Gamma^{14}_l,$ where the sites are labeled counter-clockwise and for $\hat{W}_{\square}$ the first link $\langle ij \rangle$ lies on a triangle. 
\label{fig.bandstructure}
The model is then solved by introducing a Majorana representation of the $\Gamma$-matrices using six flavors of Majorana operators, $\{\xi_{i}^{1},\xi_{i}^{2},\xi_{i}^{3},\xi_{i}^{4},c_{i},d_{i}\}$, at each site $i$: 
 \begin{equation}
 \label{eq:Gamma_Majorana}
\Gamma_{i}^{a}  =  i\xi_{i}^{a}c_{i},\quad
\Gamma_{i}^{5}  =  ic_{i}d_{i},\quad
\Gamma_{i}^{a5}  =  i\xi_{i}^{a}d_{i},
\end{equation}
where the Majorana fermions have the important property that $(\xi_i^a)^\dagger=\xi_i^a, c_i^\dagger=c_i$, and $d_i^\dagger=d_i$. The Hamiltonian written in terms of these operators then reduces to that of two species of Majorana fermions, $c$ and $d$, which are non-interacting and move in a background $\mathbb{Z}_2$ gauge field $u_{ij}$,
\begin{eqnarray}
\label{eqn:H_uij}
\mathcal{\tilde{H}} &&= J_{\triangle} \sum_{\langle ij \rangle\in\triangle} i u_{ij} c_i c_j \ + \ J'_{\triangle} \sum_{\langle ij \rangle\in\triangle} i u_{ij} d_i d_j \ 
\nonumber \\
&&+ \ J_{\hexagon}  \sum_{\langle ij \rangle\in\hexagon} i u_{ij} c_i c_j + \ J'_{\hexagon} \sum_{\langle ij \rangle\in\hexagon} i u_{ij} d_i d_j \ \nonumber\\ &&- \ J_5 \sum_i i c_i d_i.
\end{eqnarray}
Here the fields $u_{ij}$ are defined by $u_{ij}=-i\xi_{i}^{1}\xi_{j}^{2}  \;\text{if}\;ij\in\triangle$ and $u_{ij}=
 -i\xi_{i}^{3}\xi_{j}^{4}  \;\text{\text{if}\;\ensuremath{ij\in\hexagon}}$. The $u_{ij}$ are in general quantum fields with eigenvalues $\pm 1$ since $u_{ij}^2=1$, hence their identification with a $\mathbb{Z}_2$ gauge theory. Moreover, we can simultaneously diagonalize $\mathcal{\tilde{H}}$ and $\{u_{ij}\}$ since they commute. We emphasize that the interacting Hamiltonian \eqref{eqn:Hamiltonian} with the representation \eqref{eq:Gamma_Majorana} is reduced to the effectively {\em non-interacting} Hamiltonian \eqref{eqn:H_uij} because the $u_{ij}$ behave as constants in each flux sector of the $\hat{W}_\triangle$, $\hat{W}_\square$, and $\hat{W}_{\hexagon}$.
 
However, the full Hilbert space spanned by the Majorana fermions is overcomplete, and a constraint must be enforced to ensure that the Clifford algebra of $\Gamma$ operators is satisfied. This constraint is expressed by the operator equation $D_{i}=-\Gamma_i^1 \Gamma_i^2\Gamma_i^3\Gamma_i^4\Gamma_i^5 =-i\xi_{i}^{1}\xi_{i}^{2}\xi_{i}^{3} \xi_{i}^{4}c_{i}d_{i}=1$ and is enforced by the projector $P=\prod_{i}\left[\frac{1+D_{i}}{2}\right]$, where the product is over all sites in the lattice. The original Hamiltonian is obtained  through ${\cal H}=P {\cal \tilde H}P$.

The $Z_2$ gauge fields define fluxes $\phi_p$ via $\exp(i\phi_p)\equiv \prod_{jk\in p} iu_{jk}$, where $jk$ is taken counterclockwise on each elementary plaquette $p$. These fluxes are then related to the conserved $\hat{W}$'s by $W_{p} \propto \prod_{ij\in p}u_{ij}$, where $ij$ is also taken counterclockwise. Since the eigenstates of the Hamiltonian are also eigenstates of the fluxes, once the fluxes have been specified, the $u_{ij}=\pm 1$'s are uniquely specified up to $\mathbb{Z}_2$ gauge transformations, and the Hamiltonian $\mathcal{\tilde{H}}$ is then diagonalized and projected to yield to eigenstates of $\mathcal{H}$. The ground state is then determined by a many-body Majorana wavefunction minimizing the total energy. Due to translational symmetry, $\mathcal{\tilde{H}}$ describes a bandstructure for the $c$ and $d$ fermions.  It can be shown that the minimal energy configuration is the one where all the negative ``eigenstates'' of the effective band Hamiltonian \eqref{eqn:H_uij} are occupied.\cite{Kitaev:ap06,Yao:prl09,Wu:prb09}

One finds that $\phi_p=\pm \pi/2$ in the triangular plaquettes, and $\phi_p=0,\pi$ in the hexagonal and square plaquettes. Under time reversal symmetry, $W_p\to \pm W_p$ where $-(+)$ is for triangle (hexagon and square) plaquettes; it follows that $\phi_p \to -\phi_p$ for triangle plaquettes while $\phi_p$ remains unchanged for hexagon and square plaquettes. Consequently, a ground state with a certain flux pattern $\{\phi_p\}$ spontaneously breaks time reversal symmetry. The ground state energy of a flux configuration must be degenerate with the flux pattern obtained from $\{\phi_p\}$ by changing $\phi_p\to-\phi_p$ on all triangular plaquettes.\cite{Kitaev:ap06,Yao:prl09} 

To determine the physical ground state, the flux configuration which minimizes the ground state energy needs to be determined. This was accomplished by first numerically determining the ground state energy for each flux configuration for the symmetric couplings  $J_\triangle=J'_{\triangle}=J_{\hexagon}=J'_{\hexagon}=J$ and $J_5=0$. The ground state flux with the least energy is depicted by Fig.\ref{fig:lattice_ruby}(b). Generally the ground state flux will be a function of $J_5$, but for simplicity we will not consider additional flux configurations in this work because an additional term (depending on $\hat{W}_\triangle,\hat{W}_{\square},\hat{W}_{\hexagon}$) can always be included to favor a particular configuration without destroying the exact solvability.\cite{Tikhonov:prl10,Chua:prb11} By general arguments regarding the Majorana representation of the spin operators in the 3/2 representation,\cite{Kitaev:ap06,Baskaran:prl07} the spin-spin correlation functions of the ground state are identically zero beyond nearest neighbor sites. Hence, the ground state is also a quantum spin-liquid in these observables. 

With the gauge sector of $\mathcal{H}$ specified, we can solve our system by mapping the Majorana fermions to complex fermions $a_i$, defined by $c_i = a_i + a^\dagger_i$ and $d_i = -i (a_i - a^\dagger_i)$. This puts our Hamiltonian into the form

\begin{eqnarray}
\label{eqn:H_ucd}
\mathcal{\tilde{H}} = \sum_{\langle ij \rangle} &&\left\{ i (u^c_{ij} + u^d_{ij}) a^\dagger_i a_j + \frac{1}{2} i (u^c_{ij} - u^d_{ij}) (a^\dagger_i a^\dagger_j + a_i a_j) \right\}
\nonumber \\
&&\qquad \qquad \qquad \qquad- \ 2 \hspace{.4mm} J_5 \sum_i (a^\dagger_i a_i - \frac{1}{2}),
\end{eqnarray}
where we have defined the quantities $i u^c_{ij} = i u_{ij} J_{\triangle (\hexagon)}$ and $i u^d_{ij} = i u_{ij} J'_{\triangle (\hexagon)}$ for $ij \in \triangle (\hexagon)$. This type of Hamiltonian is solved by taking a Bogoliubov transformation in momentum space. Taking the Fourier transform of \eqref{eqn:H_ucd} and introducing the Nambu spinor as $\Psi^\dagger (k) = \left( a_1 (k) \ \ a_2 (k) \ \ \cdots \ \ a_6 (k) \ \ a^\dagger_1 (-k) \ \ \cdots \ \ a^\dagger_6 (-k) \right)$ where $a_i(k)$ is the Fourier mode of the fermion creation operator on the $i$th site in the unit cell, we obtain

\begin{eqnarray}
\label{eqn:H_bdg}
\mathcal{\tilde{H}} = \sum_{k \in C_+} \Psi^\dagger (k) H(k) \Psi (k),
\end{eqnarray}
with the identifications

\begin{eqnarray}
\label{eqn:hdelt}
H(k) &&=
\left( \begin{matrix}
h(k)&\Delta(k)\\
\Delta(k)&-h^T(-k)
\end{matrix}\right),
\nonumber \\
h_{ij}(k) &&= i \left(\tilde{u}_{ij}^c(k) + \tilde{u}_{ij}^d(k)\right) + 2 \hspace{.4mm} J_5 \delta_{ij},
\nonumber \\
\Delta_{ij}(k) &&= \frac{1}{2} i \left(\tilde{u}^c_{ij}(k) - \tilde{u}^d_{ij}(k)\right),
\end{eqnarray}
where the $\tilde{u}$'s are the Fourier modes of the corresponding $u$'s defined above, and where $C_+$ is half of the Brillouin zone. We only sum over half of the Brillouin zone to avoid the double counting introduced by the Nambu spinor. Because $H(k) = -H(-k)$, the eigenvalues of $H(k)$ appear in pairs $\{E_j(k),-E_j(-k)\}$, $j = 1,2,...,6$, and our model is particle-hole symmetric. This is a consequence of our Hamiltonian being written in terms of the Nambu spinor, which satisfies $\Psi^\dagger_i(-k) = \Psi_i(k)$; when we diagonalize $H(k)$, half of the states are redundant.

\section{Band structure and Phase diagram}

\begin{figure}
\vspace{0.5cm}
\includegraphics[height=7cm]{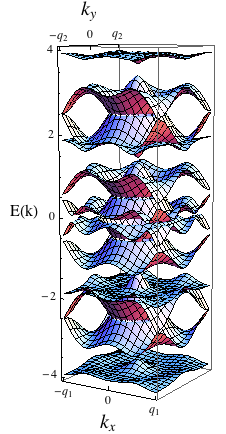}
\caption{Bandstructure for the coupling parameters $\{J_{\triangle},J'_{\triangle},J_{\hexagon},J'_{\hexagon},J_5\}$=$\{1.0,0.75,1.0,1.33,1.0\}$ where $q_1 = (2\pi/3)(\sqrt{3}-1)$ and $q_2 =\pi(1-1/\sqrt{3})$.. There are 12 bands which exhibit a redundant particle-hole symmetry and the energy spectrum is gapped with Chern number 2.} \label{fig:bandstructure}
\end{figure}

For general couplings the ground state may be gapped or gapless. For the gapped phases, the band structure may possess a non-trivial (non-zero) Chern number. We will refer to these non-trivial gapped phases as Chern phases. The physical consequence of a non-zero Chern number is the appearance of chiral gapless edge modes in a system with a boundary.  If the system has a finite Chern number, it will exhibit a quantized thermal Hall conductance. We find these signatures of a non-trivial topological phase by calculating the Chern number numerically\cite{Avron:prl51,Kells:njp11} and diagonalizing a system with boundaries (a ``strip" geometry) to determine the existence of gapless edge modes.  This is most conveniently done by taking a Fourier transform to momentum space along the ``length" of the strip geometry.\cite{Kitaev:ap06,Yao:prl09,Wu:prb09} 

\begin{figure}
\includegraphics[width=5cm]{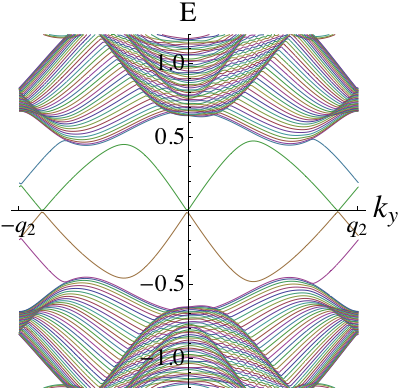}
\caption{Spectrum of the strip geometry of the system as a function of the momentum $k_y$, parallel to the edge. The system has Chern number -1 and couplings $\{J_{\triangle},J'_{\triangle},J_{\hexagon},J'_{\hexagon},J_5\}$=$\{1.0,0.6,1.0,1.67,0.2\}$. Two chiral edge modes that traverse the band gap at $E=0$ are clearly visible and each corresponds to a mode on a single edge. } \label{fig:edge}
\end{figure}

An example band structure is shown in Fig.\ref{fig:bandstructure}, which illustrates the dispersion of Majorana excitations $E(\vec k)$ as a function of the two planar momentum components $k_x,k_y$.  The state in Fig.\ref{fig:bandstructure} is gapped about zero energy (below which all states are occupied and above which all states are empty) and has a non-trivial Chern number.  There are six bands above and six bands below zero energy, corresponding to the 6 sites in the unit cell of the ruby lattice.   Shown in Fig.\ref{fig:edge} is the spectrum of a  strip geometry in the non-trivial phase with Chern number -1.  
 
\begin{figure}
\includegraphics[width=7cm]{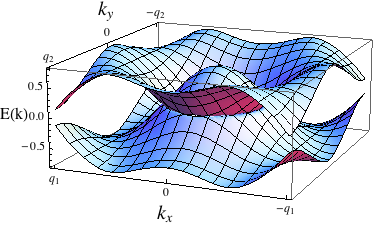}
\includegraphics[width=5cm]{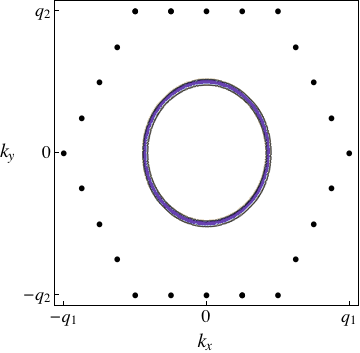}
\caption{(Color online) Bandstructure and shape of the Fermi surface (blue) at  $\{J_{\triangle},J'_{\triangle},J_{\hexagon},J'_{\hexagon},J_5\}=\{1.0,1.0,1.0,1.0,1.4\}$ and where $q_1 = (2\pi/3)(\sqrt{3}-1)$ and $q_2 =\pi(1-1/\sqrt{3})$. The Fermi surface pocket is located around Brillouin zone center. The dots are indicative of the zone boundary.} \label{fig:rubysurf1}
\end{figure}

\begin{figure}
\includegraphics[width=7cm]{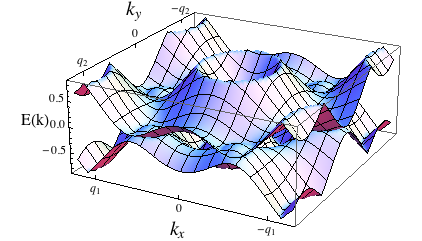}
\includegraphics[width=5cm]{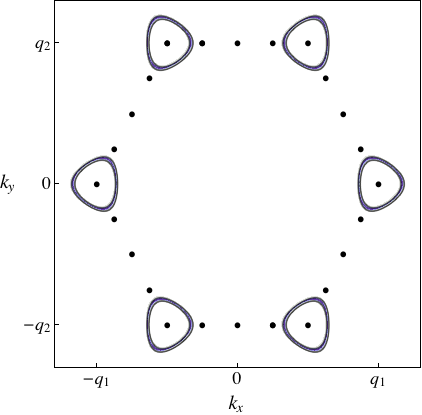}
\caption{(Color online) Bandstructure and shape of the Fermi surface (blue) at  $\{J_{\triangle},J'_{\triangle},J_{\hexagon},J'_{\hexagon},J_5\}=\{1.0,1.0,1.0,1.0,0.55\}$ and where $q_1 = (2\pi/3)(\sqrt{3}-1)$ and $q_2 =\pi(1-1/\sqrt{3})$. The Fermi surface pockets are located around the $K$ and $K'$ symmetry points of the Brillouin zone. The dots are indicative of the zone boundary.} \label{fig:rubysurf2}
\end{figure}

\begin{figure}
\includegraphics[width=7cm]{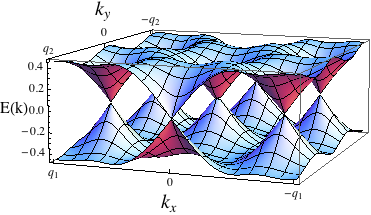}
\caption{Bandstructure of a nodal phase with four nodes and at parameters $\{J_{\triangle},J'_{\triangle},J_{\hexagon},J'_{\hexagon},J_5\}=\{1.0,0.58,1.0,1.73,2.62\}$. The nodes are located at the zone center and the three M points of the Brillouin zone.} \label{fig:node}
\end{figure}

When the ground state is gapless, it generically has a Fermi surface. Shown in Figs. \ref{fig:rubysurf1} and \ref{fig:rubysurf2} are band structures with a Fermi surface realized in the ground state.  In Fig.\ref{fig:rubysurf1} the Fermi surface is in the center of the first Brillouin zone (BZ), while in Fig.\ref{fig:rubysurf2} the Fermi surface is near the high symmetry $K$ and $K'$ points of the hexagonal BZ.  Note that the Fermi surface is ``connected" once the appropriate reciprocal lattice vectors are used to shift them together.  At a phase boundary between two gapped phases with different Chern numbers, the ground state is gapless with Dirac nodes at a discrete set of points in the BZ. An example of such a band structure is shown in Fig.\ref{fig:node}, which has four nodes in the first BZ (one at the zone center and three others at the inversion symmetric $M$ points).   

\begin{figure*}
\includegraphics[width=15cm]{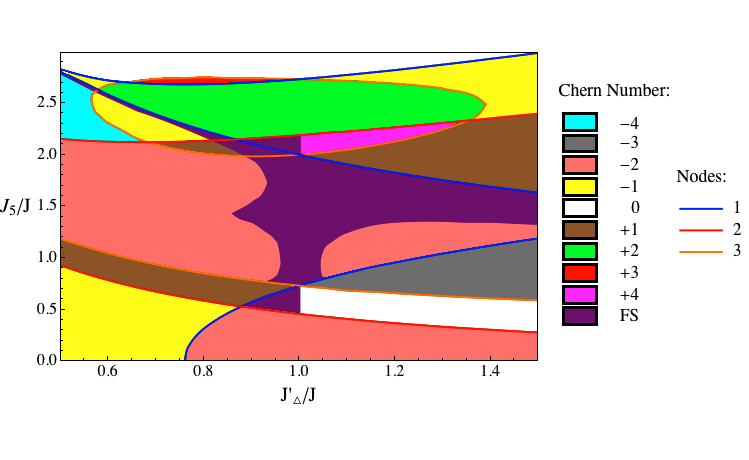}
\caption{(Color online) Phase diagram of the spin-liquid at couplings $J_{\triangle}=J_{\hexagon}=J$ and $J'_{\triangle}\times J'_{\hexagon}=J^2$. The different colored regions correspond to either a gapped phase with Chern number as given in the legend or if gapless correspond to a Fermi surface (FS) phase. Lines which are boundaries between gapped phases are gapless phases with nodes or Fermi points at specific points in the Brillouin zone. The single node phase is gapless at the zone center $\mathbf{k} = (0,0)$ and can be either a Dirac cone or a quadratic band touching point; the double node phase is gapless at the $K$ and $K'$ points with $\mathbf{k} = (\pm \pi(\sqrt{3}-1)/3,\pm \pi (1-1/\sqrt{3}))$, and the triple node phase is gapless at the three inversion symmetric $M$ points, $\mathbf{k} = (0,\pm \pi (1-1/\sqrt{3}))$ and $\mathbf{k} = (\pm \pi(\sqrt{3}-1)/2,\pm \pi (1-1/\sqrt{3})/2)$. These lines are labeled according to the number of such nodes.}\label{fig:phasediagram}
\end{figure*}

In general, the phase diagram is complicated due to the large number of tunable parameters, even when one insists on translational, six-fold rotation and inversion symmetry. We report on a phase diagram resulting from an interesting selection of parameters which reveals the general richness of the model. We explored the phase diagram for the specific case where the couplings $J_{\triangle}=J_{\hexagon}$ are held equal to a fixed constant $J$, and the $J'$ couplings are allowed to vary in such a way that their product remains a constant; explicitly $J'_{\hexagon} \times J'_{\triangle}=J^2$.  This phase diagram is shown in Fig.\ref{fig:phasediagram} and is exceptionally rich with many gapped and gapless phases.

The regions of the phase diagram are classified into whether or not they are gapped or gapless, and whether or not the gapless phases possess a non-trivial Chern number (Chern phase). If gapless, the phases are classified according to whether the gap closes at discrete Fermi points in the BZ or along a line or Fermi surface (FS). Often the former contains Dirac nodes on the BZ boundary or the zone center where the Fermi point may also be a quadratic band touching point. The Fermi points only occur on critical lines separating two Chern phases or a Chern phase and a Fermi surface. We will refer to these critical phases as nodal lines (NL). 

The phase diagram in Fig.\ref{fig:phasediagram} contains NLs with up to 3 Fermi points. Since the NLs are obtained by solving a secular equation of a Bloch Hamiltonian which is analytic, the NLs trace out smooth trajectories in the phase diagram. Moreover, the complicated manner in which they cross yields the many Chern phases in Fig.\ref{fig:phasediagram}. Most interestingly, we see a 3 node NL which closes in on itself to form an ellipse. 

Another interesting region is the high symmetry line at constant $J'_{\triangle}/J=1$ where $J_{\triangle}=J'_{\triangle}=J_{\hexagon}=J'_{\hexagon}=J$. This line contains an gapless FS phase in a line between $0.7\lesssim J_5 \lesssim 2.0$, and there are four NLs which cross this FS line at its endpoints. Along this line there are no nearby Chern phases in the phase diagram. This FS phase is non-critical in regards to tuning $J'_{\triangle}$ and $J'_{\hexagon}$, appears to be particularly stable, and is the FS phase with the largest volume in the phase diagram. 

Finally, we note that in the phase diagram transitions between Chern phases and FS phases involve critical NLs. This transition will lead to Fermi surface pockets forming around the nodes located at the points of high symmetry. Figures \ref{fig:rubysurf1} and \ref{fig:rubysurf2} are examples of such pockets. There are also transitions between Chern phases and FS phases which do not involve a NL. In this case the Fermi surface becomes gapped by a pairing (Cooper) instability as the system is driven into the Chern phase. 

In the next section of this paper, we will study the effects of the inclusion of new interaction terms which spoil the exact solvability of the model. These new interactions will also motivate the study of the Ising order in one the gamma matrices, namely $\Gamma_i^5$.

\section{Ising Order of $\Gamma^5$}

In this section we consider the following extension to $\mathcal{H}$,
\begin{eqnarray}
&&\mathcal{H} = \mathcal{H}_0 + \mathcal{H}_5 + \mathcal{H}_\text{Ising}, \nonumber \\
&&\mathcal{H}_{\text{Ising}} =  - \lambda \sum_{\langle ij \rangle} \Gamma^5_i \Gamma^5_j.
\end{eqnarray}
We also limit ourselves to case where $\lambda >0$. For our analysis it is convenient to take the spin-orbit interpretation of the model where we have 4 local degrees of freedom with basis states $\{ \left| \uparrow \uparrow \right\rangle, \left| \uparrow \downarrow \right\rangle, \left| \downarrow \uparrow \right\rangle , \left| \downarrow \downarrow \right\rangle\}$. In this basis, $\Gamma_i^5$ is trivially diagonal with eigenvalue $+1$ if the $1/2-$spins are aligned and eigenvalue $-1$ if they are anti-aligned. Thus, we can regard $\Gamma^5$ as an Ising spin, albeit one that is doubly degenerate for each Ising polarization. Then $\mathcal{H}_\text{Ising}$, as its name suggests, describes ferromagnetic nearest neighbor Ising-Ising interactions in the case where $\lambda >0$. It is also trivially true that $[\mathcal{H}_\text{Ising} , \Gamma_i^5 ]=0$ for all sites $i$, which implies that both operators are simultaneously diagonal in the local spin-orbit basis. Thus in the $\mathcal{H}_0 = 0$ limit, the model is trivially exactly solvable with eigenstates given by tensor products of local $\Gamma_i^5$ eigenstates. A typical eigenstate would have the form,
\begin{eqnarray}
&&\left|\psi \right\rangle = \begin{array}{c}\\\bigotimes\\ \scriptstyle{i\in I}\end{array}\left( \alpha_i \left|\uparrow\uparrow\right\rangle +\beta_i \left|\downarrow\downarrow\right\rangle \right)\begin{array}{c}\\\bigotimes\\ \scriptstyle{j\in J}\end{array}\left( \gamma_j \left|\uparrow\downarrow\right\rangle +\delta_j \left|\downarrow\uparrow\right\rangle \right),\nonumber \\
&&|\alpha_i|^2 + |\beta_i|^2 = |\gamma_j|^2 + |\delta_j|^2 = 1,  
\end{eqnarray}
where the $N$ sites of the lattice are partitioned into disjoint sets $I$ and $J$ with $N_I$ and $N_J$ sites each respectively. The energy of such a state is then given by
\begin{eqnarray}
\left( \mathcal{H}_5 + \mathcal{H}_\text{Ising} \right)\left|\psi \right\rangle  = \left\{ (N_I - N_J )J_5 -\lambda (N_\text{link} - L_\text{dom.})\right\}\left|\psi \right\rangle,  \nonumber \\ \nonumber \\
\end{eqnarray}
where $N_\text{link}$ is the total number of nearest neighbor links and $L_\text{dom.}$ is the total length of all the boundaries between the different ``Ising domains". The arbitrariness in the parameters $\alpha, \beta, \gamma, \delta$ on every site leads to a macroscopic degeneracy of the state $\left|\psi\right\rangle$; this stems from the local degeneracy of $\Gamma^5_i$. Also, an exact eigenstate of $\Gamma_i^5$ will exhibit no fluctuations in the $\Gamma_i^5$ observables. Hence, the ground states of $\mathcal{H}_\text{Ising} + \mathcal{H}_5$ are \emph{trivial} Ising ferromagnets with order in $\langle \Gamma_i^5\rangle=\pm 1$. The sign of the order parameter will depend on the sign of $J_5$ if it is non-zero. Otherwise, if $J_5=0$, the ground state will spontaneously break the Ising symmetry in $\Gamma_i^5$. 

However, when $\mathcal{H}_0 \neq 0 $ the situation is no longer trivial and no longer exactly solvable. In particular, $[\mathcal{H}_0 , \Gamma^5_i ] \neq0$ suggests that it would be difficult to diagonalize the Hamiltonian in the spin-orbit basis or any basis where $\Gamma_i^5$ is diagonal. Nevertheless, one still has $[W_p,\mathcal{H}_\text{Ising}]=0$ for all plaquettes $p$. This implies that the flux invariants are still good quantum numbers for both the GMM and $\mathcal{H}_\text{Ising}$, and it is sensible to solve for the eigenstates within a given flux sector $\{W_p\}$. Writing $\mathcal{H}_\text{Ising}$ after performing the Majorana transformation makes this explicit,
\begin{equation}\label{eqn:Ising_Majorana}
\tilde{\mathcal{H}}_\text{Ising} = -4\lambda \sum_{\langle ij \rangle} a_i^\dagger a_i a_j^\dagger a_j + 2\lambda\sum_i a_i^\dagger a_i - 2N\lambda.
\end{equation}  
Hence, $\tilde{\mathcal{H}}_\text{Ising}$ does not couple to the $\mathbb{Z}_2$ gauge field degrees of freedom and the fluxes are conserved. If we now consider again the limit where $\mathcal{H}_0=0$, then $\tilde{\mathcal{H}}$ describes a trivial insulating Hamiltonian with no kinetic terms but with local $a$-number conservation. One can then write down the exact eigenstates in this representation which is just any occupation state of the complex fermion $a_i$ for all sites $i$ where the flux configuration can remain arbitrary. This gives another explanation of the macroscopic degeneracy of $|\psi\rangle$, which in the gauge field picture is due to the many degenerate $\{W_p\}$ sectors. Equivalently, the flux invariant operators are insensitive to whether or not the local 1/2-spins are aligned or anti-aligned; rather, they operate on degrees of freedom that are orthogonal. We can now also give a physical interpretation to the complex fermion $a$. Namely, its occupation parity $P_i= (2a_i^\dagger a_i-1)$ is the $\Gamma_i^5$ observable whose eigenvalue determines whether or not the local 1/2-spin are aligned or anti-aligned. Also, the $a$ occupation numbers do not fluctuate in an eigenstate, which is consistent with the non-fluctuating behavior of the Ising order parameter. 

Now we return to considering general $\mathcal{H}$. Note that $\mathcal{H}_0$ neither commutes with $\Gamma_i^5$ nor its sum total $\sum_i\Gamma_i^5$. This leads to the non-conservation of the total $a$-particle number in the ground state, and we can associate this non-conservation with the pairing terms in the $a$-representation of the problem. Nevertheless, the product of parities as defined by $\mathcal{P}=\prod_i P_i = \prod_i \Gamma_i^5$ is conserved since $\mathcal{H}_0,\mathcal{H}_5$ and $\mathcal{H}_\text{Ising}$ all individually commute with it. In fact $\mathcal{P}$ implements a global $\mathbb{Z}_2$ gauge transform, $a_i\leftarrow -a_i,\quad a_i^\dagger \leftarrow -a_i^\dagger$ on every site $i$. This is merely another manifestation of the conservation of the total $a$-number modulo $2$ by $\mathcal{H}$. 

Next we will consider the opposite limit where $\mathcal{H}_\text{Ising} + \mathcal{H}_5 = 0$ and argue that the exact ground states of $\mathcal{H}_0$ will not possess any Ising order in $\Gamma^5$ by themselves, but that the exact ground states of the more general GMM with $J_5\neq 0$ will. Moreover, in the limit of small $J_5$, these ground states are adiabatically connected. Given a fixed flux sector $\{W_p\}$, and given the existence of a non-degenerate ground state of $\tilde{\mathcal{H}}_0$ which we denote by $|\Omega\rangle$, we will show that $\langle\Omega|\Gamma_i^5|\Omega\rangle = 0$ for all $i$.

First consider equation (\ref{eqn:H_ucd}) in the limit $J_5=0$ where the gauge fields $u_{ij}$ have been fixed. Define a linear \emph{unitary} particle-hole conjugation operator $\mathcal{C}$ by its by conjugation on $a_i$ and $a_i^\dagger$ and its action on the $a$-number vacuum $|0\rangle$ as follows:
\begin{eqnarray}
\mathcal{C}a_i\mathcal{C}^\dagger = a_i^\dagger, \quad \mathcal{C}a_i^\dagger\mathcal{C}^\dagger = a_i, \nonumber \\
\mathcal{C}|0\rangle = \mathrm{e}^{i\theta}a_i^\dagger \ldots a_N^\dagger |0\rangle.
\end{eqnarray}
$\theta$ is a phase which we can fix by demanding that $\mathcal{C}^2|0\rangle =|0\rangle$. For $N$ even this leads to $\theta=\frac{N\pi}{4}$. These relations then totally specify $\mathcal{C}$ and also imply the relations $\mathcal{C}^\dagger=\mathcal{C}$ and $\mathcal{C}^2=1$. Note that $[\tilde{\mathcal{H}}_0,\mathcal{C}] =0$, implying that if $|\Omega\rangle$ is non-degenerate, then $\mathcal{C}|\Omega\rangle$ can only differ from $|\Omega\rangle$ by a phase. Thus, $\langle \Omega |a_i^\dagger a_i | \Omega \rangle = \langle\Omega|\mathcal{C}^\dagger a_i a_i^\dagger \mathcal{C}|\Omega\rangle = \langle\Omega|a_ia_i^\dagger|\Omega\rangle$. Using $\{a_i,a_i^\dagger\}=1$, we then conclude that $\langle\Omega|a_i^\dagger a_i|\Omega\rangle=1/2$ and $\langle\Omega|\Gamma_i^5|\Omega\rangle=0$. Since $a^\dagger_i a_i$ is $\mathbb{Z}_2$ gauge invariant, we expect this to hold true even after projection. One could also heuristically argue that because $J_5$ couples to $a^\dagger_i a_i$ like a chemical potential, the ground state will have a particle-hole symmetry and is the half-filled Fermi sea. However, the presence of pairing terms $a_i a_j +a^\dagger_i a^\dagger_j$ complicates this argument.

For the model that was solved in Sec. II where $J_5 \neq 0$ generally, $\langle \Gamma_i^5 \rangle$ may take non-zero values in $(0,1]$ when in the ground state. In this more general situation, the presence of $\mathcal{H}_5$ breaks the symmetry under $\mathcal{C}$ conjugation. In addition, since the ground states of the GMM were shown not to undergo a phase transition as $J_5$ is tuned from zero, we conclude that the exactly solvable ground states of the GMM may exhibit Ising order in the $\Gamma_i^5$ spins. That is, the exactly solvable phase and Ising order are \emph{not mutually exclusive}. Hence the ground state is \emph{not} a spin-liquid in these observables. But fluctuations in $\Gamma_i^5$ remain non-trivial in the exactly solvable phase. This poses the interesting question as to what extent an exactly solvable GMM ground state is adiabatically connected to the trivial Ising states such as the one described by $|\psi \rangle$ above. 

In the limit where $\mathcal{H}_0$ is treated as a perturbation to $\mathcal{H}_5+\mathcal{H}_\text{Ising}$, the weakly fluctuating Ising order $\Gamma_i^5$ will strongly renormalize the $J_5$ coupling as seen by the $a$ fermions, which can be seen from equation (\ref{eqn:Ising_Majorana}). Thus, we can expect this limit to be equivalent to a large $J_5$ limit and to be described by a low energy effective Hamiltonian derived from perturbation theory. Schematically, the Hamiltonian can written in terms of $W_p$ operators,\cite{Kitaev:ap06,Kells:prl08,Yao:prl09} 
\begin{equation}
\mathcal{H}_\text{eff} = \sum_p \alpha_p W_p + \ldots.
\end{equation}
where $\{\alpha_p\}$ are coupling constants and the dots represent higher order terms. Such an effective theory will break the degeneracy among the different flux sectors and favor certain flux sectors over others. However, time reversal symmetry is still preserved and must be spontaneously broken by the ground state. If $J_5=0$, then the ground must also spontaneously break the Ising symmetry of $\Gamma^5$. 

In the opposite limit where $\mathcal{H}_\text{Ising}$ is the perturbation to $\mathcal{H}_0+\mathcal{H}_5$, the situation is more complicated as the ground state of the exactly solvable phase is in general non-trivial. We try to address this question in the next section of the paper where $\mathcal{H}_\text{Ising}$ is treated as a perturbation to the exactly solvable GMM within the mean-field approximation.  

\begin{figure}
\includegraphics[width=\columnwidth]{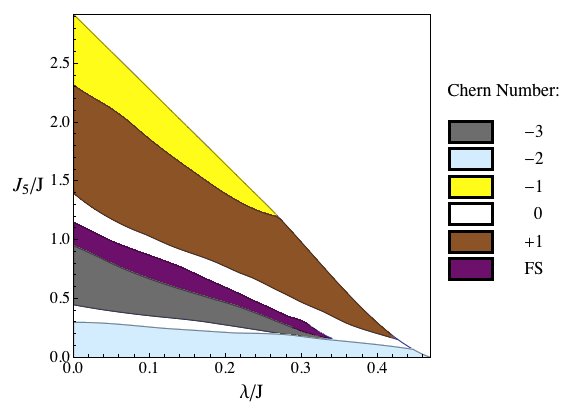}
\caption{(Color online) Phase diagram with the Ising perturbation treated within the mean field approximation. The couplings are $\{J_{\triangle},J'_{\triangle},J_{\hexagon},J'_{\hexagon}\}=\{1.67,1.0,0.3,1.0\}$.}
\label{fig:mfphase1} 
\end{figure}

\begin{figure}
\includegraphics[width=\columnwidth]{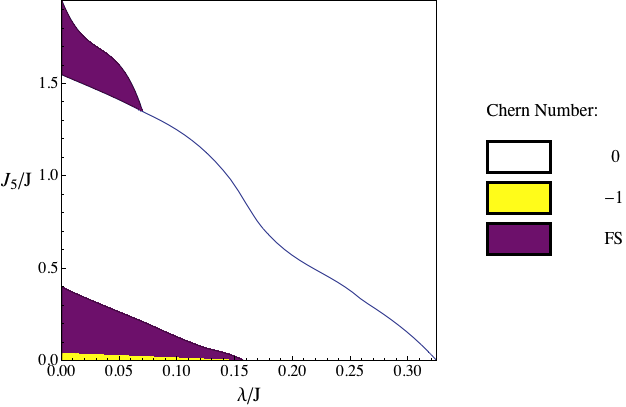}
\caption{(Color online) Phase diagram with the Ising perturbation treated within the mean field approximation. The couplings are $\{J_{\triangle},J'_{\triangle},J_{\hexagon},J'_{\hexagon}\}=\{1.0,1.0,0.2,0.2\}$. The phase boundary separating the two trivial phases is a first order transition line where $\langle \Gamma_i^5 \rangle$ changes discontinuously.}\label{fig:mfphase2} 
\end{figure}

\section{Mean Field Approximation of the Ising Perturbation}

In this last section we study the effects of $\mathcal{H}_\text{Ising}$ as a perturbation to the exactly solvable $\mathcal{H}_0+\mathcal{H}_5$ within the mean-field approximation. Performing a mean-field decomposition with respect to the order parameter $\langle \Gamma^5_i\rangle$ results in the following mean-field Hamiltonian,
\begin{equation}
\mathcal{H}_{\text{Ising}} \approx \mathcal{H}_{\text{MF}} = - 4 \lambda \sum_{\langle ij \rangle}  \langle \Gamma^5_i \rangle (a_i^\dagger a_i -1/2 ) + \lambda \sum_{\langle ij \rangle} \langle \Gamma^5_i \rangle \langle \Gamma^5_j \rangle,
\end{equation}
where the mean field $\langle \Gamma^5_i\rangle$ has to be determined self-consistently. Motivated by the fact that the Ising couplings are ferromagnetic, we make the ansatz that $\langle \Gamma^5_i\rangle$ is uniform across the entire lattice. At the mean-field level, the effect of the interaction $\mathcal{H}_\text{Ising}$ can already be seen to renormalize $J_5$ with $J_5^\text{eff}=J_5 + 2 \lambda \langle \Gamma^5 \rangle$.  

Within this approximation, we determined the phase diagrams for varying $J_5$ and $\lambda$ with fixed $J$ and $J'$ couplings and fixed flux sector. Two phase diagrams are shown in Fig.\ref{fig:mfphase1} and Fig.\ref{fig:mfphase2}. We note that generically the topological phases are stable against weak $\mathcal{H}_\text{Ising}$ perturbations. At larger $\lambda$, however, the energetics always prefer the trivial gapped phase which has zero Chern number and large $\langle \Gamma^5 \rangle$. Hence, large $\lambda$ eventually renormalizes $J_5$ to larger effective values. This is consistent with the large $J_5$ and large $\lambda$ trivial phases being adiabatically connected to each other. 

Focusing along the line at constant $\lambda=0$ and increasing $J_5$, we see a series of gapped and gapless phases which eventually ends with a trivial phase at large $J_5$. This is expected from the results of the previous section, but since $J_5$ acts like an external field, the Ising order is $\langle \Gamma^5_i\rangle \neq 0$ on this line. When $\lambda$ is tuned to greater values, the volume of these intermediate phases diminishes continuously, eventually giving way to the trivial phase. This confirms that the trivial Ising phase is \emph{adiabatically connected} to the trivial large $J_5$ phase. 

In Fig.\ref{fig:mfphase1}, the gapped phase at $J_5=0$ appears to be the most stable with respect to increasing $\lambda$. However, for the couplings shown in Fig.\ref{fig:mfphase2}, the $J_5=0$ phase gives way to a trivial gapped phase (Chern number zero) which is not adiabatically connected to the trivial phase at large $J_5$. This phase undergoes a first order phase transition to the larger $J_5$ trivial phase as shown in Fig.\ref{fig:mfphase2}. Thus, the Ising interaction may choose to stabilize certain gapped phases over the default $J_5=\lambda=0$ phase.   The stabilized phases are highly dependent on the particular values of the other coupling constants.    

At constant $J_5>0$, increasing $\lambda$ may stabilize certain phases which would typically require larger $J_5$ values. For example, in Fig.\ref{fig:mfphase1} along the line $J_5=0.75$, the system is driven to a Fermi-surface phase with increasing $\lambda$. Other examples of such transitions driven by the Ising interaction may also been seen in Figs. \ref{fig:mfphase1} and \ref{fig:mfphase2}. Hence, the Ising perturbation treated at mean-field may stabilize certain exotic phases which are otherwise only accessible by intermediate values $J_5$. 

In summary, these results show that with increasing $\lambda$, the ground state is eventually adiabatically connected to the trivial Ising ordered state of Sec. IV. However, in general, various phase transitions involving other exotic phases must occur before this happens. Moreover, this depends crucially on the initial ($\lambda =0$) exactly solvable phase. 

\section{Conclusions}
In this work we studied an exactly solvable $\Gamma$-matrix generalization of Kitaev's original spin-1/2 model on the ruby lattice shown in Fig.\ref{fig:lattice_ruby}.  Our model can be interpreted as describing either a spin-3/2 system or a double-orbital spin-1/2 model.  We have shown that the ruby lattice, with its many possible tunable parameters, exhibits a rich phase diagram  with many exotic phases realized by same Hamiltonian. We have found gapless phases with Fermi surfaces and Fermi points, as well as gapped topologically non-trivial phases. Next, we studied the effects of including Ising-like interaction terms that destroy the exact solvability of the model. We analyzed the ordered phase that these terms lead to and argued that the ground state in the large $J_5$ limit of the exactly solvable model is adiabatically connected to it. We then analyzed in detail the general model near the limit of an exactly solvable phase and treat the Ising terms as perturbations within the mean-field approximation.  We derived phase diagrams which show that the Ising interactions will eventually favor the trivial Ising ordered phase of the model over the other more exotic phases. We also confirm that the large $J_5$ exactly solvable phase is adiabatically connected to the trivial Ising ordered phase. 

While this study has focused on a particular exactly solvable spin model and special perturbations on it, connections previously drawn to other topological phases indicate that its results are rather broadly applicable.\cite{Fiete:PhysE12}   
Moreover, a number of future directions for further study suggest themselves:  (i) Considering anti-ferromagnetic Ising interactions ($\lambda < 0 $) to determine if the interplay between the exactly solvable GMM and the Ising interactions might be different.  (ii) The effect of disorder in interacting many-body quantum systems remains a key open problem.  
The model we study here with its unusually rich phase diagram provides an opportunity to study disorder effects within a well-controlled model.\cite{Chua_dis:prb11}  (iii) Doping magnetic systems is one route to high-temperature superconductivity.  To date, no work has been reported on the doping of Gamma matrix generalizations of the Kitaev model, and only a few studies exist on the original model.\cite{You:arxiv11}  The richness of the model we study here could shed light on doping magnetic systems in a much more general context and could ultimately help guide the discovery of strongly correlated materials exhibiting high temperature or unconventional superconductivity.

\acknowledgments
We gratefully acknowledge funding from ARO grant W911NF-09-1-0527 and NSF grant DMR-0955778.

\bibliography{Kitaev}

\end{document}